\documentclass[%
reprint,
superscriptaddress,
 amsmath,amssymb,
]{revtex4-2}

\usepackage{graphicx}
\usepackage{dcolumn}
\usepackage{bm}
\usepackage[table,x11names]{xcolor}
\usepackage[draft,inline,nomargin]{fixme}

\fxsetup{theme=color,mode=multiuser}

\FXRegisterAuthor{mh}{amh}{\color{blue}MH}
\FXRegisterAuthor{bh}{abh}{\color{red}BH}

\begin{document}

\preprint{APS/123-QED}

\title{Three-Dimensional Neutron Far-Field Tomography of a Bulk Skyrmion Lattice}
  
\author{M. E. Henderson}
\email{mehenderson@uwaterloo.ca}%
\affiliation{Institute for Quantum Computing, University of Waterloo, Waterloo, ON, Canada, N2L3G1}
\affiliation{Department of Physics \& Astronomy, University of Waterloo,
  Waterloo, ON, Canada, N2L3G1}
\noaffiliation

\author{B. Heacock}
\thanks{M.E.H. and B.H. contributed equally}
\affiliation{National Institute of Standards and Technology, Gaithersburg, Maryland 20899, USA}

\author{M. Bleuel}
\affiliation{National Institute of Standards and Technology, Gaithersburg, Maryland 20899, USA}
\author{D. G. Cory}
\affiliation{Institute for Quantum Computing, University of Waterloo, Waterloo, ON, Canada, N2L3G1}
\affiliation{Department of Chemistry, University of Waterloo, Waterloo, ON, Canada, N2L3G1}
\author{C. Heikes}
\affiliation{National Institute of Standards and Technology, Gaithersburg, Maryland 20899, USA}
\author{M. G. Huber}
\affiliation{National Institute of Standards and Technology, Gaithersburg, Maryland 20899, USA}
\author{J. Krzywon}
\affiliation{National Institute of Standards and Technology, Gaithersburg, Maryland 20899, USA}
\author{O. Nahman-Levesqu\'{e}}
\affiliation{Institute for Quantum Computing, University of Waterloo, Waterloo, ON, Canada, N2L3G1}
\affiliation{Department of Physics \& Astronomy, University of Waterloo,
  Waterloo, ON, Canada, N2L3G1}
\author{G. M. Luke}
\affiliation{Department of Physics and Astronomy, McMaster University, Hamilton, ON, Canada, L8S 4M1}
\affiliation{Brockhouse Institute for Materials Research, Hamilton, ON, Canada, L8S 4M1}
\author{M. Pula}
\affiliation{Department of Physics and Astronomy, McMaster University, Hamilton, ON, Canada, L8S 4M1}
\author{D. Sarenac}
\affiliation{Institute for Quantum Computing, University of Waterloo, Waterloo, ON, Canada, N2L3G1}
\author{K. Zhernenkov}
\affiliation{Institute for Quantum Computing, University of Waterloo, Waterloo, ON, Canada, N2L3G1}
\affiliation{J\"ulich Centre for Neutron Science at Heinz Maier-Leibnitz Zentrum, Forschungszentrum J\"ulich GmbH, 85748 Garching, Germany}
\author{D. A. Pushin}
\email{dmitry.pushin@uwaterloo.ca}
\affiliation{Institute for Quantum Computing, University of Waterloo, Waterloo, ON, Canada, N2L3G1}
\affiliation{Department of Physics \& Astronomy, University of Waterloo,
  Waterloo, ON, Canada, N2L3G1}





\date{\today}
\begin{abstract}

\end{abstract}

\maketitle

Skyrmions are topologically-protected spin textures thought to nucleate and annihilate on points of vanishing magnetization, called Bloch points \cite{milde2013unwinding, kagawa2017current,Yu2020real}. However, owing to a lack of bulk techniques, experimental visualizations of skyrmion lattices and their stabilization through defects in three-dimensions remain elusive. Here, we present a tomographic algorithm applied to a Co$_8$Zn$_8$Mn$_4$ skyrmion lattice host, processing multi-projection small angle neutron scattering measurements to generate mean scattering feature reconstructions (MSFR) of the bulk spin textures. Digital phantoms validated the algorithm; reconstructions of the sample show a disordered skyrmion lattice with a topological saturation of 63~\%, exhibiting three-dimensional topological transitions through two different emergent (anti)monopole defect pathways with densities of 147~$\mu$m$^{-3}$ and 21~$\mu \mathrm{m}^{-3}$ for branching and segmentation events, respectively. Our techniques produce experimentally-informed visualizations of bulk skyrmion lattice structures and defects, enabling future bulk studies over a wide variety of sample shapes and chemistries, magnetic phases, and external parameters.

Magnetic skyrmions manifest as spin-vortices, whose topological protection drives particle-like properties and condensation into thermodynamically stable phases in an external magnetic field \cite{nagaosa2013topological,muhlbauer2009skyrmion,tokunaga2015new,adams2011long}. Most commonly observed as 2D structures in metallic materials \cite{thinfilm,Wiesendanger2016nanoscale, Kiselev2011chiral, Yu2011near, Yu2010real, Heinze2011spontaneous}, their quantized emergent magnetic flux \cite{Skyrme1962unified,nagaosa2013topological,Schulz2012emergent,milde2013unwinding} which characterizes the Berry phase an electron accrues when following the local magnetization adiabatically \cite{schutte2014dynamics}, generates unique transport phenomena \cite{Neubauer2009topological,Jiang2017direct,Kanazawa2015discretized}, multiferroic behavior \cite{Seki2012observation}, and electric controllability via ultra-low current densities \cite{Jonietz2010spin,Schulz2012emergent}. Such a culmination of features sees skyrmions as promising candidates for next-generation low-power spintronic information-processing and storage devices \cite{Zhang2015magnetic,Sampaio2013nucleation,Tomasello2014strategy}.

In three-dimensions, the uniform stacking of the 2D skyrmion spin structure produces skyrmion tubes elongated along the external magnetic field direction, thought to penetrate surface-to-surface \cite{vanderlaan2021depth}. In physical bulk crystal systems at non-zero temperature, a finite density of defects exist, interrupting the skyrmion string propagation \cite{Iwasaki2013current}. Because the emergent flux that defines skyrmions is quantized, their nucleation and termination is mediated by emergent magnetic charges that must also be quantized \cite{milde2013unwinding}. Skrymion tube segmentation and branching via emergent magnetic monopoles and antimonopoles (denoted S$^{+}$, S$^{-}$, B$^{+}$, and B$^{-}$, respectively) are believed to mediate skyrmion topological transitions \cite{Birch2021Topological}. Motion of such defects in response to changes in external parameters, such as field or temperature conditions, have been proposed to drive a change in skyrmion topology through in the unwinding of individual skyrmions \cite{Iwasaki2013current,Birch2021Topological} and the zipping/unzipping of neighboring skyrmion tubes \cite{milde2013unwinding, kagawa2017current,Birch2021Topological}. Since total emergent charge is conserved, skyrmionic transitions can only take place in three-dimensions when emergent (anti)monopoles are either pinned to a material defect or jammed in place and unable to overcome the activation energy required to travel to the material surface or reach an oppositely-charged monopole to annihilate  \cite{thinfilm,kagawa2017current,birch2020real}.

\begin{figure*}
\includegraphics[width = \textwidth]{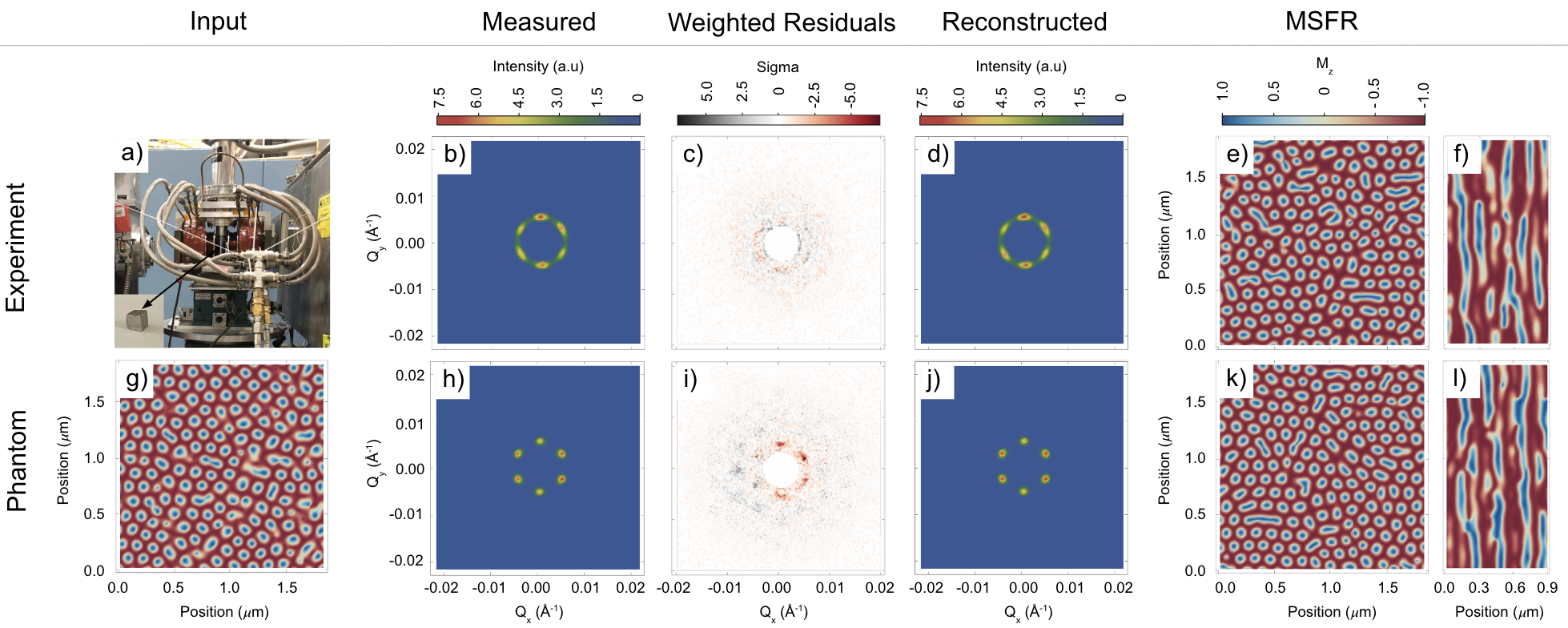}
\caption{Experimental setup (a), measured SANS image (b) reproduced from \cite{henderson2021characterization}, minimized weighted residuals (c), reconstructed SANS image (d), and xy (e) and xz (f) slices of the MSFR are shown in the upper panel. The lower panel shows the corresponding plots for the phantoms where (g) is the simulated input magnetization, (h) is the simulation SANS pattern, and (i)-(l) follow the same as the upper experimental row. All SANS images are for the projection with the guide field aligned to the neutron beam.}
\label{fig:SANS}
\end{figure*}

 Examinations of skyrmions using 2D imaging have revealed a myriad of in-plane skyrmion string deformations including elongated structures \cite{milde2013unwinding} and bent skyrmion strings which terminate on the surface \cite{Yu2020real} or form at edges \cite{birch2020real}, while 3D imaging in confined systems has revealed axial modulation of skyrmion tubes \cite{Wolf2022}. In all of these instances, the physical systems being probed are subject to constricted geometries by virtue of the thin-plate or needle-shaped samples required by the techniques. This introduces confinement effects, causing the system to exhibit drastically different energetics to those of bulk samples, favoring surface-pinning \cite{koshibae2019dynamics} and edge-interactions \cite{Du2015edge,Meynell2014surface,birch2020real} which affect skyrmion formation, shape, interactions, dynamics, and stabilization. A recent study \cite{seki2021direct} confirmed the presence of interrupted and merging-type skyrmion strings in a sparsely populated micrometer sized thin plate needle-shaped sample of Mn$_{1.4}$Pt$_{0.9}$Pd$_{0.4}$Sn using scalar magnetic X-ray tomography. The observations, however, are limited to individual skyrmion strings in a sample thickness only a few times the skyrmion tube diameters, whose confined geometry and thickness gradient fundamentally alters the skyrmions shape and behaviour; bulk lattice skyrmion behavior has yet to be experimentally observed.

Small-angle neutron scattering (SANS) is sensitive to bulk scattering features, enabling studies of truly bulk skyrmionic systems along different paths through the sample's phase diagram. In particular, spatially averaged depth information may be obtained by forming rocking curves, integrating over one or more diffraction peaks as a function of sample angle, to extract the longitudinal correlation length of a skyrmion lattice \cite{adams2011long}. The effects of recent external magnetic field and temperature history on jammed states has been shown using SANS, where skyrmion lattice defect densities were reduced using an ordering sequence where the external magnetic field is rocked relative to the sample \cite{gilbert2019precipitating}.

Whereas previous studies are confined to surface-level techniques, confined systems, or integral far-field SANS measurements which cannot produce real space representations of the sample, here we perform tomography of the thermal equilibrium triangular skyrmion lattice phase using multi-projection SANS measurements coupled with a free energy regularization to generate a three-dimensional mean scattering feature reconstruction of a bulk skyrmion lattice. The reconstruction algorithm consists of first forming an estimator of a multi-projection set of SANS measurements by operating on the incoming neutron state with a forward operator, which takes the MSFR volume as its main input.  Next, the sum of weighted residuals between the estimated and measured SANS patterns for multiple projections is minimized with respect to the MSFR. Without integrating over peak areas before forming the rocking curve, the $\chi^2$ is sensitive to shifting peak locations, shapes, intensities, and correlations. The number of free parameters depends on the chosen MSFR volume, but will usually be larger than the number of data points in the set of SANS images.  The large degeneracy of possible solutions and danger of overfitting the data is overcome by adding a free energy regularizing functional to the objective function $f = \chi^2 + \beta F$, where the $\chi^2$ is the weighted sum of measurement residuals, $F$ is the free energy of the MSFR, and $\beta$ is a Lagrange multiplier reminiscent of a Boltzmann factor.  The free energy includes the Heisenberg exchange, Dzyaloshinskii-Moriya (DM) exchange, and external field Zeeman terms and could be made to include additional interactions which are functionals of the spin density, though such terms are beyond the scope of the present work. Similar techniques are often used in traditional computed tomography (CT) algorithms, in which case the total variation can used as a regularizing functional \cite{sidky2008image}. The resulting MSFRs may be interpreted as containing the types of structures, and their densities, which are common within the sample. However, there is no portion of the sample which looks exactly like a MSFR, and there are a large number of possible MSFRs which would converge on a minimum of the objective function. One can also view minimizing the objective function as performing a micromagnetic simulation with the $\chi^2$ providing the local interaction and pinning potential terms in the free energy that cause lattice defects, thereby enforcing the lattice correlation lengths and structure encoded in the SANS patterns. 

 \begin{figure*}
\includegraphics[width = \textwidth]{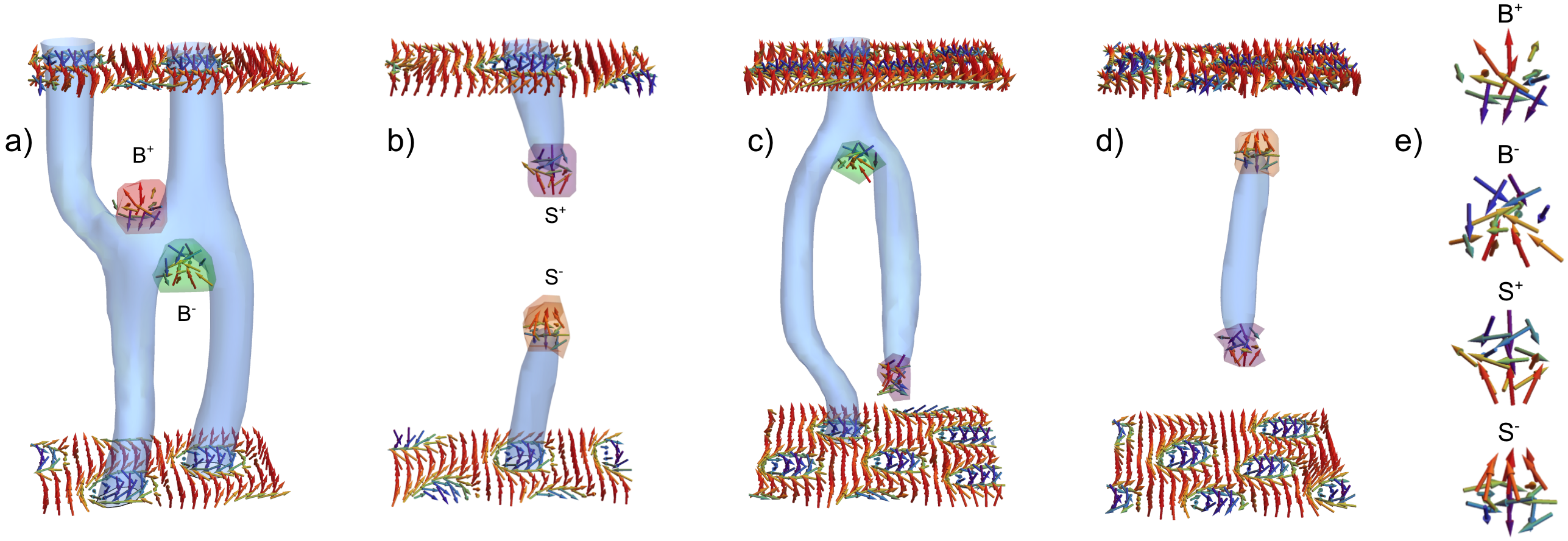}
\caption{Characteristic topological transitions present in the MSFR showing branching and segmentation (anti)monopole pathways. The blue contours outline the skyrmion tubes where the out of plane magnetization, $m_{z} = -0.5 $. Some of the skyrmion tubes are masked to highlight the regions undergoing changes in topology. Emergent magnetic charge density contours are shown for the two types of skyrmion topological transition processes with S$^{+}$, S$^{-}$, B$^{+}$, and B$^{-}$ corresponding to purple, orange, red, and green, respectively. Branching (B) and segmentation (S) emergent (anti)monopoles are observed to occur at transition points along the tubes, displaying distinct spin-textures determined by the sign of $\partial_z m_{z}$. Zoomed in spin textures are shown for the labeled branching and segmentation (anti)monopoles with each spin corresponding to one voxel (e). All skyrmion features are shown for a depth of 37 voxels, equating to 531~nm (see materials and methods).  }
\label{fig:3D}
\end{figure*}

Multi-projection SANS measurements were made using an above room temperature skyrmion host $\mathrm{Co}_{8}\mathrm{Zn}_{8}\mathrm{Mn}_{4}$ previously characterized in \cite{henderson2021characterization}.  This class of materials has been studied across a variety of techniques including SANS \cite{tokunaga2015new,Karube2016robust,henderson2021characterization,Karube2020Metastable,prebinger2021vital, SkyrmionHenderson}, magnetic susceptibility \cite{Karube2020Metastable,tokunaga2015new,Karube2016robust,henderson2021characterization}, and Lorentz Transmission Electron Microscopy (LTEM) \cite{tokunaga2015new,Karube2016robust}. The sample and magnetic field were rotated through 30 projections over a 5.8~degree angular range, which was limited by the geometry of the guide field coils (Fig. 1a).  
To determine the implications for how a limited-projection dataset would effect our results, we reconstructed digital phantoms under both ideal conditions, where SANS data was simulated for 31 projections between $\pm 15$~deg about both axes (no compound rotations) and the actual experimental projections (see methods, supplementary materials).

\begin{table*}
\setlength\tabcolsep{0.1in}
\renewcommand{\arraystretch}{1.5}
\begin{tabular}{|l|c|c|c|c|c|c|c|c|c|}
\hline
     & $h$ & $\langle m_z \rangle_\mathrm{seed} $ & $\langle m_z \rangle$  & $N_\mathrm{sk} / N_\mathrm{max}$ & 
    $\rho_\mathrm{branch}, \; \mu \mathrm{m}^{-3} $ & $\rho_\mathrm{seg}, \; \mu \mathrm{m}^{-3}$ & $\rho_\mathrm{tot}, \; \mu \mathrm{m}^{-3}$  \\ \hline \hline
  Phantom A   & 0.56 & 0.40 & 0.33   & 65~\%   & 101   & 37   & 138   \\ \hline
  MSFR A, ideal  & 0.50   & 0.30 & 0.28    & 52~\%   & 322   & 7  & 329 \\  \rowcolor{PaleGreen3} \cellcolor{white} &  0.60   & 0.35 & 0.32    & 59~\%   & 238   & 13  & 251 \\  & 0.75   & 0.40 & 0.37    & 65~\%   & 150   & 26  & 176 \\ 
  \hline
  MSFR A, lim. proj. & 0.50   & 0.30 & 0.30    & 56~\%   & 299   & 9  & 308 \\ \rowcolor{PaleGreen3}  \cellcolor{white} &0.60   & 0.35 & 0.34   & 60~\%   & 233   & 13  & 246\\ & 0.75   & 0.40 & 0.40    & 64~\%   & 164   & 24  & 189    \\ \hline \hline 
  Phantom B   & 0.75 &  0.40 & 0.41  &  68~\% & 47 & 40 & 87  \\ \hline
    MSFR B, ideal  & 0.60  & 0.35 & 0.35  & 62~\%  & 138  & 10  &  148 \\ 
                    & 0.75  & 0.40 & 0.38  & 64~\%  & 115  & 12  &  127 \\ 
                 \rowcolor{PaleGreen3} \cellcolor{white} & 0.87  &  0.45 & 0.42 & 72~\%  & 53  &  31 & 84  \\ \hline
MSFR B, lim. proj.  &  0.60  & 0.35 & 0.35  & 60~\% &  166 & 6  &  172  \\ \rowcolor{PaleGreen3} \cellcolor{white}
                       & 0.75  &  0.40 & 0.41 & 67~\% & 88  & 16  &  104  \\ 
                     &  0.87  & 0.45 & 0.45  & 69~\% & 53  &  31 &  84  \\ \hline \hline 
  Phantom C   & 0.94 & 0.40 & 0.47   & 73~\%   & 13  & 103  & 115   \\ \hline
  MSFR C, ideal & 0.80   & 0.42 & 0.41    & 69~\%   & 133   & 27  & 161\\ \rowcolor{PaleGreen3} \cellcolor{white} & 0.94   & 0.47 & 0.46   & 72~\%   & 54  & 56  & 110\\ 
   \rowcolor{PaleGreen3} \cellcolor{white} & 1.08   & 0.52 & 0.49    & 72~\%   & 21   & 91  & 112     \\ \hline
  MSFR C, lim. proj. & 0.80   & 0.42 & 0.43    & 69~\%   & 126   & 26  & 152\\ \rowcolor{PaleGreen3} \cellcolor{white} & 0.94   & 0.47 & 0.49  & 72~\%  & 46   & 63   & 109\\
                & 1.08   & 0.52 & 0.54    & 71~\%   & 23   & 95  & 119    \\ \hline \hline 
  MSFR, sample  &  0.50 &  0.30 & 0.25  &  53~\% &  214 & 6  &  220  \\ 
                &  0.60 &  0.35 & 0.35 &  59~\% &  163 & 13 & 176 \\ \rowcolor{PaleGreen3} \cellcolor{white}
                &  0.75 &  0.40 & 0.42  &  63~\% & 118  &  21 & 139 \\
                &  0.87 &  0.45 & 0.48  &  66~\% &  69 & 37  &  106  \\ \hline
\end{tabular}
\caption{Zeeman term weight in reduced field units $h$, seeded and final average magnetization, topological saturation, and defect densities of the phantoms and associated MSFRs and sample MSFR.  The reduced field is fixed; all other parameters are computed from the spin density.}
\label{tab:metas}
\end{table*}

Two-dimensional slices of the measured and reconstructed SANS images, weighted residuals, and MSFRs are shown for the bulk sample and phantom set B with average magnetizations of 0.42 and 0.41, respectively. (Fig.~\ref{fig:SANS}). Videos showing magnetization and topological defects as a function of depth for the phantoms and MSFRs are available with the supplementary materials. Average magnetization, topological saturation, and defect densities are compared in Table~1. Phantoms were generated with the external field near the helical-skyrmion boundary (Phantom A), the ideal value (Phantom B), and the ferromagnetic-skyrmion boundary (Phantom C). The required weighting of the Zeeman term $h$, to achieve the average magnetization that was estimated from DC-susceptibility, was affected by the $\chi^2$. The favored MSFR are thus those with values of $h$ which produce the estimated average magnetization and are highlighted in Table~1. The agreement between phantom and MSFR is reasonable for Phantoms B and C. However, the MSFR for Phantom~A overestimates the branching defect density. This is likely attributable to the seeding, where Phantom A transitioned from an average magnetization of $\langle m_z \rangle =0.4$ to $\langle m_z \rangle =0.33$ during the free energy relaxation (see Materials and Methods), while the MSFR had a net magnetization change of $\langle m_z 
\rangle =0.35$ to $\langle m_z \rangle =0.32$, suggesting hysteresis-like effects can impact the fidelity of the reconstructions. 

The sample MSFR topological saturation $N_\mathrm{sk} / N_\mathrm{max} $ is found to be $63~\%$ of that of a perfect, skyrmion triangular lattice with the same $Q_0$. This is reflected by the presence of transverse and longitudinal lattice distortions. In particular, skyrmion voids, bimerons, and elongated spin structures present in the two-dimensional MSFR magnetization slices reduce the number of skyrmions from that of an ideal hexagonally packed lattice. Similarly, interruption of the skyrmion strings along their length may be visualized by taking an $x-z$ slice of the MSFR as shown in Fig. 1f.

Three-dimensional visualizations of portions of the MSFRs show skyrmion nucleation and annihilation along their depth, where segmentation S$^\pm$ and branching B$^\pm$ (anti)monopoles mediate the change in topology at skyrmion transition points (Fig. 2e). These features are similar to those observed for simulations of skyrmion annihilation in three-dimensions through helical and conical pathways \cite{Birch2021Topological,milde2013unwinding,kagawa2017current}. Notably, some branching events are observed to occur along wave vectors $Q$ which are offset by 60 degrees from the horizontal nearest-neighboring skyrmion, producing a skyrmion twisting effect (Fig.~2a). Conversely, some instances of segmentation events exhibit pairs of S$^\pm$ (anti)monopoles which cup skyrmions, producing spatially localized skyrmion filaments which extend longitudinally over a few lattice periods (Fig.~2d) and are reminiscent of magnetic torons \cite{toron}. Given the field history of the sample, wherein saturation to the ferromagnetic phase was performed prior to collection of the tomography data, these structures may represent a skyrmion survival mechanism in the field-polarized state via magnetic torons on defect pinning sites \cite{SkyrmionHenderson}. Future studies will be performed to examine the prevalence of these structures as a function of field history and average magnetization, shedding insight into skyrmion elongation and stabilization mechanisms.

The energy source required for emergent (anti)monopole creation may derive from internal chemical disorder present in the material (such as in material imperfections or site-disorder \cite{Birch2021Topological,Yu2010real,schutte2014dynamics}), thermal activation \cite{Oike2016Interplay,Magnetic2021Tokura,Inplane2021Mathur}, and the external magnetic field setting relative to the helical and ferromagnetic phase boundaries \cite{Magnetic2021Tokura,Yu2020real,milde2013unwinding}. 
For the case of the magnetic field setting, segmentation and branching (anti)monopoles are thought to control skyrmion annihilation upon increasing and decreasing fields, respectively \cite{milde2013unwinding,kagawa2017current,Zheng2018experimental}.  This is reinforced by reconstructions performed on equivalent datasets under different average magnetization and reduced field $h$ conditions; a shifting prevalence from branching defects in the low-field case, to segmenting defects in the high-field case is observed.  

The preferred sample MSFR shows segmentation and branching (anti)monopoles seemingly jammed in place and unable to travel to the sample surfaces or annihilate. The observed defect densities could be due to pinning potentials in the material which would trap the (anti)monopoles and inhibit their motion \cite{schutte2014dynamics}. In this case, the prevalence of these structures may indicate the degree of internal chemical disorder, providing snapshots of magnetic defect pinning centers. Alternatively, these features may be evidence of an incomplete ordering sequence, wherein the limited magnetic field directions during rotation did not allow the monopoles to propagate along enough angular paths in the sample  to completely breakup the jammed labyrinth domains \cite{gilbert2019precipitating}. Future SANS tomography measurements taken as a function of skyrmion ordering could decouple these two possibilities. Implementation with structured neutron beams, which possess an analogous non-trivial winding character in spin \cite{sarenac2018generation} or phase \cite{Dima2015,Sarenac2022}, may provide estimates of the magnetic defect densities based on the vertical widening of the skyrmion peaks in the transverse geometry. Alternatively, the defects themselves may be viewed as the magnetic equivalent of phase singularities, capable of generating topological neutron states for probing nanometric sample topologies.

We have shown experimentally-informed visualizations of a bulk triangular skyrmion lattice, uncovering 3D topological transitions which clearly exhibit a mixture of distinctive segmenting and branching (anti)monopole defects. Unique features for these two event types are observed in the MSFR, characterized by skyrmion twisting during branching events and localized skyrmion filament structures cupped by S$^\pm$ (anti)monopoles. Our SANS tomography techniques open the door to future studies of bulk micromagnetic materials on unprecedented lengthscales, including skyrmion and emergent (anti)monopole structure, behaviour, and dynamic processes.

\begin{acknowledgments}
BH would like the thank Dustin Gilbert for suggesting that skyrmion systems would be an interesting subject to study with SANS tomography. This work was supported by the Canadian Excellence Research Chairs (CERC) program, the Natural Sciences and Engineering Council of Canada (NSERC) Discovery program, the Canada First Research Excellence Fund (CFREF), and the National Institute of Standards and Technology (NIST) and the US Department of Energy, Office of Nuclear Physics, under Interagency Agreement
89243019SSC000025. Access to SANS and CHRNS was provided by the Center for High Resolution Neutron Scattering, a partnership between NIST and the National Science Foundation under Agreement No. DMR-1508249. 

\end{acknowledgments}

\bibliography{references}

\section*{materials and methods}

The $\mathrm{Co}_{8}\mathrm{Zn}_{8}\mathrm{Mn}_{4}$ material was grown via a modified Bridgman technique in a rapid high temperature furnace at McMaster University, cut into a cube of dimensions 3.4 mm x 3.3 mm x 3.0 mm, and oriented such that the (100) direction was coming out of the major face. Details of the sample, as well as single-projection SANS and susceptibility measurements are available elsewhere \cite{henderson2021characterization}. 

Multi-projection, unpolarized SANS was performed at the NG7-30m beamline at the National Institute for Standards and Technology (NIST) for a 15 m beam configuration at a wavelength of $6$~\AA. The sample was field cooled from 420 K in a field of 250 G to 310 K and the external field was manually rocked back and forth to achieve the observed six-fold pattern \cite{SkyrmionHenderson}.

Digital phantoms were made by seeding an LLG relaxation using Ubermag \cite{beg2021} with a high-energy lattice made to have a fixed correlation volume. The LLG relaxation was stopped after ten iterations to retain a reasonable density of defects. The system used the free energy functional

\begin{equation}
    E[\pmb{m}] = -A\pmb{m} \cdot {\nabla}^2\pmb{m} +D\pmb{m} \cdot (\nabla \times \pmb{m}) -\mu_{0}M_{s} \pmb{H} \cdot \pmb{m} .  
\end{equation}

\noindent
with exchange stiffness $A=10$~pJ/m, Dzyaloshinskii–Moriya constant  $D = 3.93 \times 10^{-3}$~J/m, saturation magnetization $M_{s}=1\times 10^{6}$~A/m, and external field values of $H =$ 0.2080~T, 0.2780~T, and 0.3475~T for Phantoms A, B, and C, respectively.  

The seed was generated by alternating between Fourier-space and real-space constraints. The magnitude of the sample magnetization in Fourier-space is set by desired correlation volume; the real space magnitude of the sample magnetization was constrained to unity everywhere; and the transverse magnetization was set to be in the direction of the curl of the longitudinal magnetization. 

SANS patterns were simulated by applying a forward operator to a randomly-selected and randomly-translated one of twenty phantoms made with differing random initialization. The resulting SANS images from repeating this process at least 100 times were averaged to simulate an incoherent neutron source and create the resulting simulated multi-projection SANS data.

The voxel size of the MSFR is determined by the Fourier-space span of the SANS images $dx = 2 \pi / Q_\mathrm{tot} = 2 \pi / (dQ \; N)$. In this experiment $dx = 14.3 \; \mathrm{nm}$, as determined by the resolution $dQ = 3.4 \; \mu \mathrm{m} ^{-1}$ and size $N \times N = 128 \times 128$ of the SANS images.  The height of the MSFR was set to 256 voxels, for a total volume of 1.8~$\mu$m$\times$1.8~$\mu$m$\times$3.7~$\mu$m$=12.3 \; \mu$m$^{3}$.

To compute the reconstructions, the cost function was minimized using a conjugate gradient method. The $\pmb{m}^2 = 1$ constraint was enforced by defining search directions in terms of the angular fields $\Theta = \arcsin (m_z) $ and $\Phi = \arctan(m_y,m_x)$. Ten iterations were run with equal weights for all residuals, which can aid in convergence when measurement uncertainties are dominated by counting statistics and there are regions of low count rates. Following the ten iterations, 100 iterations were run with the weights given the measurement uncertainties provided by the SANS reduction software \cite{kline2006reduction}. A maximum weight was introduced for the low-count rate regions to aid in convergence. The resulting average magnetization was found to depend on the weight given to the Zeeman term $h$, so reconstructions over a range of $h$ and starting average magnetization $\langle m_z \rangle $ were performed (Table~1), with the preferred MSFR being the one that most closely matches the phantom or experimental average magnetization. 

Details on the forward propagator, gradient, and defect density calculations are provided in the Supplementary Materials.

\section*{Supplementary Materials}

All MSFRs and phantoms listed in Table 1 have accompanying videos that show magnetization and defects as a function of depth which can be found here: \cite{}.

\subsection*{Forward Propagator}

The scattering pattern from passing through a single MSFR volume $S_{Q \theta}$ for some projection $\theta$ is computed by propagating a neutron wave function through the sample via the time-evolution operator $\mathcal{U}$

\begin{equation}
    S_{Q \theta} = \left | \Big \langle Q \middle | \mathcal{U} \middle | K_\theta \middle \rangle \right |^2 .
\end{equation}

\noindent
Which is estimated using a translation operator along the $z$-direction

\begin{equation}
    \mathcal{J}_{\theta, dz} = e^{i K_z (\theta, Q) dz}
\end{equation}

\noindent
assuming elastic scattering 

\begin{equation}
    K_z(\theta, Q) = \sqrt{K^2 - (K \sin \theta - Q_x )^2 - Q_y^2} ,
\end{equation}

\noindent
where $K$ is the incoming neutron wave number, along with potential the potential $V_z$ at layer $z$

\begin{equation}
    \mathcal{U}_{\theta, z + dz} = \mathcal{J}_{\theta,dz} \left [ 1 - i V_z dz \,m_n / K_z  \right ] ,
\end{equation}

\noindent
where $m_n$ is the neutron mass, and $dz$ is the voxel height. The potential from the sample is computed from its resulting magnetic field by applying the characteristic $- \hat{Q} \times \hat{Q} \times$ operator in Fourier-space \cite{sears1989neutron}

\begin{equation}
\pmb{B}_s = - 4 \pi M_s \mathcal{F}^{-1} \left \{  \hat{Q} \times \hat{Q} \times \mathcal{F} \left [  \pmb{m} \right ] \right \}
\end{equation}

\noindent
where $M_s$ is the saturated magnetization, with $4 \pi M_s = 1900$~G estimated from DC susceptibility measurements taken at 310~K.  The potential also includes the external field and operates on a neutron spinor via Pauli matrices

\begin{equation}
    V = - \mu_n \pmb{\sigma} \cdot (\pmb{B}_s + \pmb{H}_\mathrm{ext}).
\end{equation} 

\noindent
The unpolarized cross section 

\begin{equation}
    S_{Q,\theta} = \sum_{s,s'} \left | \langle Q,s' | \mathcal{U} | K_\theta, s \rangle \right |^2
\end{equation}

\noindent is then computed by summing over input $s$ and selected $s'$ spin states. 

Assuming the longitudinal correlation length of the sample is smaller than the reconstruction height, the observed scattering pattern can be estimated by self-convolving the scattering pattern of a single MSFR $N$ times

\begin{equation}
    I_{Q \theta} = \sum_{Q'',Q'} Y_{QQ'} W^\theta_{Q'Q''} \left [ \left ( S * \right )^N I_0 \right ]_{Q'' \theta} ,
\label{eqn:IQt}
\end{equation}

\noindent
where $N$ is the ratio of the sample dimension along the propagation direction and the corresponding size of the MSFR. For this sample $N = 860$. Because the forward propagator is a function of the projection of the momentum transfer in the $x-y$ plane $Q_{xy}$, the sparse matrix $W$ shifts the scale of $Q$ to be horizontal to the propagation axis. The sparse matrix $Y$ smears out $Q = K \Omega_Q \simeq K_\mathrm{avg} \Omega_Q (1 - \delta \lambda / \lambda ) $, where $\Omega_Q$ is the scattering angle, over the neutron's incoming wavelength distribution.

\subsection*{Seeding}

The initial guess is formed by first estimating the three-dimensional vector amplitude of the MSFR in Fourier-space $|\widetilde{m}_0|^2$. This is accomplished by first performing a $N^\mathrm{th}$-order deconvolution of the scattering signal with itself and the incoming beam profile, which solves for $I_\theta$

\begin{equation}
    I_\mathrm{meas} = (I_\theta *)^N I_0 
\end{equation}

\noindent
with the $I_\theta$ slices of $|\widetilde{m}|$ for each measured projection unknown. This is accomplished with modified Richardson-Lucy algorithm, which is iterative

\begin{equation}
    I_{\theta, i+1} = I_i \left [ \left (\frac{I_\mathrm{meas}}{\epsilon + (I_{\theta, i} *)^N I_0}   \right ) * (I_{\theta, i} *)^{N-1} I_0 \right ]
\end{equation}

\noindent
where $\epsilon$ is a small regulator. When $N$ is large, this can be unstable, and we found it is often helpful to enact this algorithm in stages, with $I_\mathrm{meas}$ deconvolved from $I_0$, then performing the deconvolution in $n$ stages with $N' = N^{1/n}$. It can also be helpful to track the error of the deconvolution and stop iterating, or change the regulator size when the error stops decreasing with iteration number.

After deconvolution is complete, the Bragg peaks were fit to a Lorentzian profile with respect to $Q_z$. The projection with the incoming beam aligned with the dominant direction of the skyrmion tubes $I_{\theta=0}$ was then taken as the $|\widetilde{m}|$-slice at zero $Q_z$ and expanded along the $Q_0 = D/J$ sphere, but attenuated according to the fitted Lorentzian with respect to $Q_z$. The result is a smooth function with a Lorentzian envelope along $Q_z$ that preserves the transverse correlation structure of $I_{\theta = 0}$. All the $Q>0$ structure is then scaled and a DC term is introduced to $\widetilde{m}_0$ to generate the expected net sample magnetization $\langle m \rangle = \widetilde{m}(Q=0)$.

After forming an estimate of $|\widetilde{m}_0|$, a guess for $m$ is generated with an alternating projections algorithm. The vector field of the MSFR $m$ is iteratively amended according to its constraints in Fourier and real-space. The Fourier-space constraint is given by the magnitude estimated from the deconvolved SANS data $|\widetilde{m}_0|$

\begin{equation}
    \widetilde{\pmb{m}}_{i+1} = \widetilde{\pmb{m}}_i \frac{|\widetilde{m}_0|}{|\widetilde{m}_i|}
\end{equation}

\noindent
For the first few iterations, the transverse components of $\widetilde{m}$ are also redefined according to the sign of the DM term and expected curl around $m_z$

\begin{equation}
    \widetilde{m}_{xy} = \pm i Q_{yx} \widetilde{m}_z .
\end{equation}

\noindent
The real space constraint is $\pmb{m}^2 = 1$. However, it can also be beneficial to let $m$ relax through a few iterations of a free energy minimizer before reapplying Fourier-space constraints. The net result is a guess $m$ that adheres reasonably-well to the measured SANS projections, while also having a low free energy.

\subsection*{Minimization}

The chosen cost function was

\begin{equation}
    f = \chi^2 + \beta \mathcal{F} ,
\end{equation}

\noindent
where the $\chi^2$ is the sum of weighted residuals for all the projections

\begin{equation}
    \chi^2 = \sum_{Q,\theta} (I_{Q,\theta} - M_{Q,\theta})^2 w_{Q,\theta}
\end{equation}

\noindent
and the free energy includes symmetric and antisymmetric exchange terms and a Zeeman term

\begin{equation}
    \mathcal{F} = - \frac{1}{2} \pmb{m} \cdot \nabla^2 \pmb{m} + Q_0 \, \pmb{m} \cdot (\nabla \times \pmb{m}) - \pmb{h} \cdot \pmb{m}
\end{equation}

\noindent
where the reduced field $\pmb{h} = \pmb{H} Q_0^2 / J $, and helical shell radius $Q_0 = D/J$ is taken from the SANS patterns. The $\chi^2$ depends on the difference between the measured SANS patterns $M_{Q,\theta}$ and estimated SANS intensity for each projection $I_{Q,\theta}$ computed from the MSFR via Eqn~\ref{eqn:IQt}. The residuals are weighted according to $w_{Q,\theta} = 1/\sigma_{Q,\theta}^2$, which is taken from the SANS reduction software \cite{kline2006reduction}, with the dominant uncertainty from Poisson counting statistics.

The Lagrange multiplier $\beta$ acts like a Boltzmann factor. Since the $\chi^2 \sim \mathcal{O}(\sqrt{N_{Q,\theta}})$, where $N_{Q,\theta}$ is the total number of measured SANS pixels over all projections, and the two kinetic terms in $\mathcal{F} \sim \mathcal{O}(Q_0^2)$, a reasonable choice is $\beta = \sqrt{N_{Q,\theta}} / Q_0^2$. Adjusting the weight of the Zeeman term $h$ will change the average magnetization $\langle m_z \rangle $. A reasonable value of $h$ can be selected based on studying the behavior of the free energy term in isolation \cite{balkind2019magnetic}. However, we found that larger values of $h$ were needed when the $\chi^2$ was introduced to achieve the same average magnetization, likely because the $\chi^2$ acts like a kinetic term which further reinforces the helical shell size $Q_0$. We therefore chose to study the behavior of the reconstructions over a range of $h$, as is shown in Table~1.

The gradient of the free energy term was computed via finite difference methods, similar to OOMMF \cite{donahue1999oommf}. The derivative of the $\chi^2$ necessitates taking the derivative of the forward operator, which results in computing the overlap of the forward-propagating wavefunction $\psi_\theta$ with a backward-propagating residual wavefunction $\chi_\theta$ for each projection

\begin{widetext}
\begin{equation}
\frac{\delta}{\delta \pmb{m}} \chi^2 =  i \mu_n M_s m_n dz \
\sum_\theta \frac{1}{K_z} \; \mathcal{F}^{-1} \left \{ \mathcal{F} \left \{ \mathrm{Im} \left [ \chi^\dagger_\theta  \pmb{\sigma} \psi_\theta \right ] \middle \} \cdot \middle ( \mathbb{I} - \hat{Q} \hat{Q} \right ) \right \} ,  
\label{eqn:dXdm}
\end{equation}
\end{widetext}

\noindent
where $\mathcal{F} \{ \cdots \}$ and $\mathcal{F}^{-1} \{ \cdots \}$ indicate forward and reverse Fourier transforms, respectively. The wavefunctions are computed by a combination of forward and backward propagation operators

\begin{equation}
\begin{aligned}
    \psi_\theta &= \left \langle x \Big | \mathcal{U}_z \middle | K_\theta \right \rangle \\
    \chi_\theta &= \sum_Q \left \langle x \middle | \mathcal{U}^{-1}_{z,f} \middle | Q    \right \rangle \widetilde{\psi}_{f,Q} G_Q \\
    G_Q &= 2 \sum_{Q',Q''} Y_{Q,Q'} W_{Q'',Q'} \left \{ I_0 * \left [ (I - M) w \right ]  \right \}_{Q'',\theta}
\end{aligned}
\end{equation}

\noindent
where tildes indicate Fourier-space representations; subscript $f$ denotes the final state; and $\mathcal{U}^{-1}_{z,f}$ is the backward in time propagator, starting from the final MSFR layer.

\subsection*{Defect Densities}

The emergent magnetic field is computed from the MSFR and its derivatives

\begin{equation}
    b_i = {1 \over 2} \epsilon_{ijk} \pmb{m} \cdot \left (  \partial_j 
    \pmb{m} \times \partial_k \pmb{m} \right) ,
\end{equation}

\noindent
where $\epsilon_{ijk}$ is the fully antismmetric tensor;  $\partial_i \equiv \partial / \partial x_i$; and repeated indices are summed over the three space coordinates. Summing $z$-component of this field over an area is identified as the skyrmion winding number in the enclosed area, and the emergent magnetic charge density is defined as the source term for the emergent magnetic field

\begin{equation}
    4 \pi \rho_\mathrm{em} = \nabla \cdot \pmb{b} .
\end{equation}

\noindent
A peak-finding algorithm was used to identify local maxima of $\rho_\mathrm{em}^2$ that survive a threshold cut after a Gaussian blurring. The total emergent charge of the defect was then estimated to be the sum of $\rho_\mathrm{em}$ over the neighboring $\pm 2$ voxels in all directions. Further classification of branching (two skyrmion events) versus segmentation (single skyrmion events) was accomplished by summing $\partial_z m_z$ over the same neighborhood. The same or differing signs of the two summations indicate segmentation and branching defects, respectively.

\end{document}